\documentclass[11pt,twoside]{article}
\usepackage{./asp2010}

\resetcounters 

\begin{document}

\title{Solving the $3+1$ GRMHD equations in the eXtended Conformally Flat
Condition: the XNS code for magnetized neutron stars}
  
\author{N. Bucciantini$^{1,2}$, A.G. Pili$^3$, L. Del Zanna$^{3,1,2}$}

\affil{$^1$INAF - Osservatorio Astrofisico di Arcetri, Firenze, Italy}
\affil{$^2$INFN - Sezione di Firenze, Italy}    
\affil{$^3$Dipartimento di Fisica e Astronomia, Universit\`a di Firenze, Italy}     


\begin{abstract}
High-energy phenomena in astrophysics involve quite generally a
combination of relativistic motions and strong gravity. The
simultaneous solution of Einstein equations and General Relativistic MHD equations is
thus necessary to model with accuracy such phenomena. The so-called
\emph{Conformally Flat Condition} (CFC) allows a simplified treatment of
Einstein equations, that can be particularly efficient in those
contexts where gravitational wave emission is negligible, like
core-collapse, or the formation/evolution of  neutron
stars. We have developed a set of codes to model axisymmetric MHD
flows, in General Relativity, where the solution of Einstein equations
is achieved with a semi-spectral
scheme. Here, we will show how this framework is
particularly well suited to investigate neutron star equilibrium
models in the presence of strong magnetic fields and we will present the
XNS code, that has been recently developed and here updated to
treat poloidal and mixed configurations.
\end{abstract}

\section{Introduction}

Neutron stars (NSs) are the most compact objects in the universe
endowed with an internal structure, and they are among the most
studied objects in high energy astrophysics. NSs have  very high
surface magnetic fields, up to $10^{16}$ G for
 {\it magnetars} \citep{Mereghetti08a}. It is
this very strong magnetic field  that is responsible for most of their
phenomenology. This field is amplified during  the NS formation, 
and in principle, in the interior it could be as high as $10^{18}$
G. Strongly magnetized NSs are  at the base of the so called {\it
  millisecond magnetar} model for Long and Short Gamma Ray Bursts (GRBs)
\citep{Bucciantini_Metzger+12a,Metzger_Gioannios+11a,Bucciantini_Quataert+09a}.  
A strong magnetic field  will inevitably introduce deformations of
the NS, and this makes them ideal Gravitational Waves sources.

During formation it is reasonable
to expect that the magnetic field will rapidly settle into an
equilibrium configuration. Recently,
models for relativistic magnetized stars have  been presented either
for a purely toroidal field by
\citet[KY]{Kiuchi_Yoshida08a}, and \citet[FR]{Frieben_Rezzolla12a} or
for a purely poloidal magnetic field by
\citet{Bocquet_Bonazzola+95a}. However such
configurations are unstable 
\citep{Braithwaite_Spruit06a,Braithwaite09a}.
Stability requires a mixed configuration of toroidal and
poloidal field
usually referred as {\it Twisted Torus} (TT). TT models have been
presented so far either in Newtonian regime
\citep{Lander_Jones09a,Lander_Jones12a}, or in General
Relativity (GR) but with  a perturbative approach for the metric and/or
magnetic terms
\citep{Ciolfi_Ferrari+09a,Ciolfi_Ferrari+10a,Ciolfi_Rezzolla13a}.

The main difficulty in solving for magnetized equilibrium models in GR
is due to the complexity of Einstein equations, if an exact solution
is desired. Einstein equations reduce to a set of non-linear
 coupled elliptical partial differential equations, which can
only be solved numerically. However it is well known that non-linear
elliptical equations can be numerically unstable, depending on the way
the non-linear terms are cast. 

Here we present a novel approach to compute magnetized equilibrium models for
NSs. Instead of looking for an exact solution of Einstein equations, we
make the simplifying assumption that the \emph{Conformally Flat Condition} (CFC) 
holds for the metric. This allows us to greatly simplify the equations to be solved, and
to cast them in a form that is numerically stable (\emph{eXtended} CFC, or XCFC), 
improving upon previous perturbative works
where the metric was assumed to be spherically symmetric. This allows
us to handle  stronger fields and deformations. Results are indistinguishable from
those obtained in the correct regime. This suggests that the
simplification of our approach does not compromise the accuracy of the
results, while greatly simplifying their computation.
In this work we also describe the XNS code \citep{Bucciantini-Del_Zanna11a},
here extended to the case of poloidal fields.

\section{Spacetime metric in the Conformally  Flat Condition} 

Given a generic spacetime, 
the  line element can be written as \citep{Alcubierre08a,Gourgoulhon12a}:
\begin{equation}
ds^2 = -\alpha^2dt^2+ \gamma_{ij}(dx^i+\beta^i dt)(dx^j+\beta^j dt),
\end{equation}
where $\alpha$ is called the \emph{lapse} function, $\beta^i$ is the
\emph{shift vector}, $\gamma_{ij}$ is the \emph{three-metric}, and
$i,j=r,\theta,\phi$, if a
spherical coordinate system $x^\mu = (t, r, \theta, \phi) $ is chosen.  The assumptions
of \emph{stationarity} and \emph{axisymmetry} imply that all metric
terms are only a function of $r$ and $\theta$. A metric is said to be
\emph{conformally-flat} (CFC) if
$\gamma_{ij}=\psi^4\mathrm{diag}(1,r^2,r^2\sin^2\!\theta)$, where
$\psi$ is called the \emph{conformal factor}. 

If the energy-momentum tensor $T^{\mu\nu}$  of the matter (and fields)
distribution satisfies the \emph{circularity relations}:
\begin{equation}
t_\mu T^{\mu [ \nu} t^\kappa \phi^{\lambda ]} =0 , \quad 
\phi_\mu T^{\mu [ \nu} t^\kappa \phi^{\lambda ]} =0,
\end{equation}
where square brackets indicates antisymmetrization with respect to
enclosed indexes and we have defined $t^\mu:=(\partial_t)^\mu$, $\phi^\mu:=(\partial_\phi)^\mu$, 
then the metric is \emph{quasi-isotropic}, with
$\gamma_{ij}=\psi^4\mathrm{diag}(1,r^2,Ar^2\sin^2\!\theta)$
and $\beta^i=(0,0,\beta^\phi)$.

The  GRMHD stress-energy tensor reads
\begin{equation}
T^{\mu\nu} = (e+p+b^2) u^\mu u^\nu - b^\mu b^\nu + (p + b^2/2) g^{\mu\nu},
\label{eq:grmhd}
\end{equation}
where $e$ is the total energy density, $p$ is the pressure, $u^\mu$ is the 4-velocity
of the fluid, and $b^\mu := F^{*\mu\nu}u_\nu$ is the magnetic field as measured
in the comoving frame, and $F^{\mu\nu}$ is the Faraday tensor (the asterisk indicates
the dual).  For the static configurations assumed here $u_\mu=(-\alpha,0,0,0,)$, the
magnetic field in the lab frame $B^\mu=b^\mu$, and 
circularity holds in the case of purely toroidal [$B^\mu=(0,0,0,B^\phi)$]
or purely poloidal  [$B^\mu=(0,B^r,B^\theta, 0)$] fields. 

For mixed TT configurations, deviations from circularity become
relevant only for magnetic fields with strength $\sim 10^{19}$ G,
unrealistically high even for extreme NSs. So in general one can
safely assume circularity also in the TT case and a
\emph{quasi-isotropic} metric. However, even for highly deformed objects, i.e. for rotating NSs at the
\emph{mass shedding limit}, the value of $A$ deviates from $1$ by no
more than $10^{-4}$. A conformally flat
metric thus appears to be a good approximation, better suitable to numerical solution. 

The last assumption is that of a \emph{static} spacetime, for which $\beta^i=0$,
leading to an extra condition called \emph{maximal slicing} and to a further
simplification of the Einstein equations. In this case, the equations for the two 
remaining unknowns $\psi$ and $\alpha$ are:
\begin{equation}
\Delta \psi = [ - 2\pi \psi^6  (e + B^2/2) ] \psi^{-1} ,
\label{eq:xcfc_psi}
\end{equation}
\begin{equation}
\Delta (\alpha\psi) = [ 2\pi (\psi^6  (e + B^2/2)+\psi^6( 6p + B^2)\psi^{-2}) ]  (\alpha\psi),
\label{eq:xcfc_alpha}
\end{equation}
where $\Delta$ is the standard \emph{laplacian} operator in spherical
coordinates, and the source term have been renormalized to insure
stability, according to the XCFC approach \citep{Cordero_Carrion-Cerda_Duran+09a,Bucciantini-Del_Zanna11a}.

\section{Bernoulli Integral and Grad-Shafranov equation}

The divergence-free condition for the magnetic field allows us to rewrite the 
poloidal components as derivatives of the $\phi$ component of the
vector potential,  $A_\phi$, and the imposed symmetries will allow us
to define free functions of that quantity, defined on the so-called
magnetic surfaces.
The only  non-vanishing equation of the static GRMHD system 
is the Euler equation in the presence of an external electromagnetic field:
\begin{equation}
\partial_i p + (e \! + \! p) \,\partial_i \ln\alpha = L_i := \epsilon_{ijk} J^j B^k,
\label{eq:lorentz}
\end{equation}
where $L_i$ is the Lorentz force and $J^i = \alpha^{-1}\epsilon^{ijk}\partial_j (\alpha B_k)$
is the conduction current. Assuming a \emph{barotropic} EoS
$p=p(\rho)$, $e=e(\rho)$ ($\rho$ is the rest mass density), we find
\begin{equation}
\partial_i \ln h + \partial_i \ln\alpha  = \frac{L_i}{\rho h},
\end{equation}
where the specific enthalpy  is $h:= (e+p)/\rho$.
Integrability requires
\begin{equation}
L_i = \rho h\, \partial_i \mathcal{M} = \rho h\, \frac{d\mathcal{M}}{d
  A_\phi}\partial_i A_\phi \rightarrow
\ln{\left( \frac{h}{h_c}\right)} + \ln{\left( \frac{\alpha}{\alpha_c}\right)} - \mathcal{M}(A_\phi) = 0,
\label{eq:bernoulli}
\end{equation}
where constants are calculated at the star center.
Moreover, the $\phi$ component of the Lorentz force,
which must also vanish, implies
\begin{equation}
B_\phi = \alpha^{-1} \mathcal{I}(A_\phi).
\end{equation}
Introducing $\sigma := \alpha^2 \psi^4 r^2\sin^2\!\theta$ and $\tilde{A}_\phi := A_\phi / (r\sin\theta)$ 
and the new operator 
\begin{equation}
\tilde{\Delta}_3  \! := \!  \Delta - \frac{1}{r^2\sin^2\!\theta}  \! = \! 
\partial^2_r + \frac{2}{r}\partial_r+\frac{1}{r^2}\partial_\theta^2
+ \frac{1}{r^2\tan{\theta}}\partial_\theta - \frac{1}{r^2\sin^2\!\theta},
\end{equation}
for which $\tilde{\Delta}_3 \tilde{A}_\phi = \Delta_* A_\phi / (r\sin\theta)$
(it coincides with the $\phi$ component of the \emph{vector laplacian} in spherical coordinates),
after some calculations we retrieve the \emph{Grad-Shafranov} equation for the magnetic flux function $A_\phi$:
\begin{equation}
\tilde{\Delta}_3 \tilde{A}_\phi
+  \frac{\partial A_\phi \partial\ln (\alpha\psi^{-2})}{r \sin\theta}
+ \psi^8 r \sin\!\theta \left( \rho h \frac{d \mathcal{M}}{d  A_\phi}
+ \frac{\mathcal{I}}{\sigma}\frac{d\mathcal{I}}{dA_\phi} \right) = 0.
\label{eq:gs}
\end{equation}

\section{Numerical Scheme}

For the non-linear Poisson-like equations
Eq.~\ref{eq:xcfc_psi}-\ref{eq:xcfc_alpha} we employ a 
numerical algorithm (X-ECHO/XNS) fully presented in \citet{Bucciantini-Del_Zanna11a}, to
which the reader is referred for a complete description [see also
\citet{Del_Zanna-Zanotti+07a,Bucciantini_Del-Zanna13a} for the
treatment of the fluid MHD part]. Axisymmetric solutions are searched
in terms of a series of spherical harmonics $Y_l(\theta)$:
\begin{equation}
q(r,\theta):=\sum_{l=0}^{\infty}[A_l(r)Y_l(\theta)].
\end{equation}
The Laplacian can then be reduced to a series of radial 2nd order
boundary value ODEs for the
coefficients $A_l(r)$ of each harmonic, which are then solved using
tridiagonal matrix inversion.  This procedure is repeated until convergence, using in the
source term the value of the solution computed at the previous
iteration.

The Grad-Shafranov equation
Eq.~\ref{eq:gs} can be reduced
to the solution of a non-linear vector Poisson equation, which is
formally equivalent to the equation for the $\phi$ component of the
shift-vector in the CFC
approximation, for rotating systems. Again we use the same algorithm, with a combination of
vector spherical harmonics decomposition for the angular part, and
matrix inversion for the radial part 
\citep{Bucciantini-Del_Zanna11a}.

In all of our  models we have used 20 spherical harmonics for the
elliptic solvers and a grid in spherical coordinates in the
domain $r=[0,30]$km, $\theta=[0,\pi]$, with 250 points in the radial
direction and 100 points in the angular one.
We have verified that with these choices our results are fully
converged. 

\section{Results}

\begin{figure}[!t]
\includegraphics[bb=50 320 800 550, clip, width=14cm]{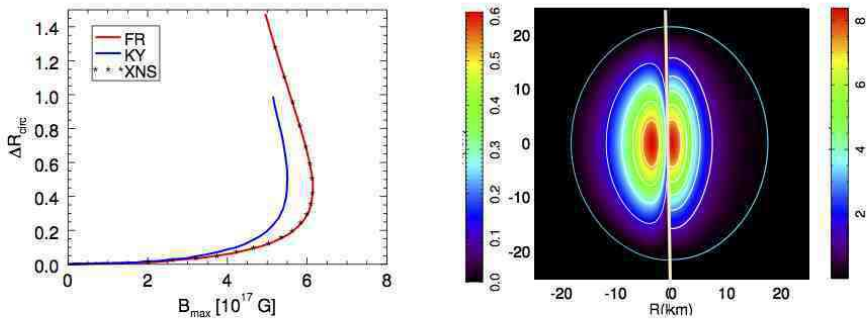}
\caption{Left: variation of the \emph{circularization radius}  (KY) as a
  function of the maximum internal magnetic field, for a NS model with
  baryonic mass $1.68 M_\odot$. Lines represent the results by FR and
  KY, points are our results with the XNS code . Right: magnetic field in units
  $10^{18}$G in the left part and density in units $10^{14}$ g
  cm$^{-3}$ in the right part. The outer line is the stellar surface.}
\label{fig:1}
\end{figure}

By choosing an appropriate form for the functions $\mathcal{I}(A_\phi),\mathcal{M}(A_\phi)$
it is possible to obtain either purely toroidal, purely poloidal, or
mixed field configurations. It is also possible to change the
structure of these configurations, or equivalently change the related
current distributions.
We have applied our XNS code and build several equilibrium sequences for
various magnetic configurations.  

An example of a purely toroidal configuration is shown in
Fig.\ref{fig:1}, where the prolate shape is evident. In
Fig.\ref{fig:1} we also show a sequence of equilibrium models, with purely
toroidal field, characterized by the same baryonic mass. Our results
are compared with what was previously found in literature (KY,FR). We confirm the more recent results by FR against the
previous one by KY. Note that FR and KY solve the same set of
equations (in the exact regime), and they both claim convergence of
the results, while we solve in the much simpler CFC. Our findings on one hand are
indicative that using a set of equations that is guarantee to be
numerically stable might be important
to assure the correctness of the results, on the other confirm that
the errors introduced by
the CFC approximation are negligible. 

\begin{figure}[!ht]
\includegraphics[bb=60 370 800 730, clip, width=14cm]{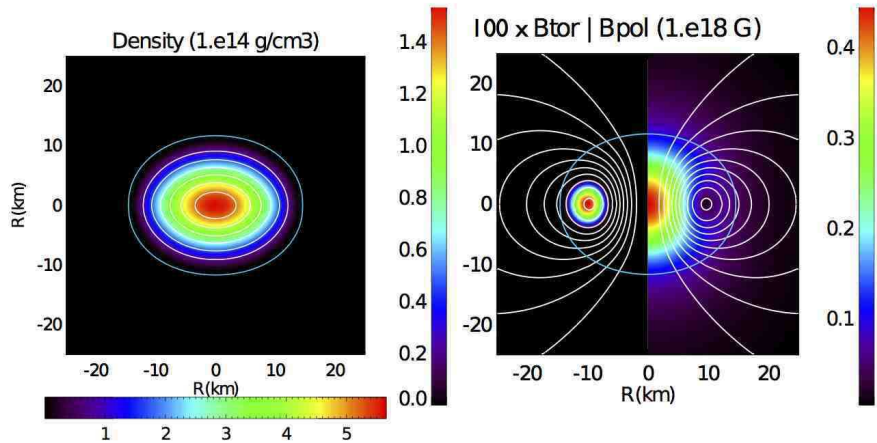}
\caption{Mixed TT configurations. Right panel: density distribution. Left
panel: strength of the toroidal (left) an poloidla (right) magneti
field components, superimposed to magnetic field surfaces. The outer
line if the stellar surface.}
\label{fig:2}
\end{figure}

For brevity we will here illustrate in detail only a TT model. In
Fig.\ref{fig:2} the magnetic field and density distribution of the TT
configuration are shown. We found that usually the toroidal field is
smaller than the poloidal one, but even when they are equal, the deformation is oblate, and
only marginally different from cases of purely poloidal
 field. This is
because the deformation is dominated by the central region, most of
the energy is in the poloidal component, and fields confined at outer radii
have marginal effects. 

To summarize:
\begin{itemize}
\item the characteristic deformation induced by a purely toroidal field, fully
  confined below the stellar surface, is prolate: the magnetic field acts
  concentrating the internal layers of the star around its symmetry
  axis, causing, on the other hand, an expansion of the outer layers;
\item given the same strength, magnetic fields concentrated in the
  outer part of the star, lead to smaller deformations, with respect
  to magnetic fields concentrated in the internal regions;
\item a purely poloidal field, that in our case extends also outside
  the star, leads to oblate equilibrium configurations: the magnetic
  stresses act preferentially in the central regions, where the field
  peaks, leading to a flatter density profile perpendicularly to the
  axis itself. We can also obtain doughnut-like configurations where the density
  maximum is not at the center;
\item the presence of additional currents located in the outer layers
  of the stars, leads only to marginal changes in its structure, and on
  the shape of the magnetic field lines outside the stellar surface;
\item for the same maximum magnetic field inside the star, purely
  poloidal configuration, are characterized by smaller deformations,
  than purely toroidal ones;
\item we have computed \emph{Twisted-Torus}
configurations in the non-perturbative regime. The toroidal component
can reach a strength comparable with the poloidal one but is
energetically subdominant. The deformation  are almost completely
due to the poloidal field, acting on the interior;
\item for a fixed central density, a higher magnetic field gives a
  higher eccentricity, a higher radius and a higher gravitational mass;
\item the more compact configurations, having a higher central
  density, can support stronger magnetic fields, and show much smaller deformations.

\end{itemize}

\bibliography{author}

\end{document}